\documentclass[
aps, 
amsmath, 
11pt,
notitlepage,
superscriptaddress
]{revtex4-1}
\usepackage{graphicx, hyperref, cleveref, xcolor}
\graphicspath{{./figures/}}

\begin{document}

\title{Helical boundary modes from synthetic spin in a plasmonic lattice}

\author{Sang Hyun Park}
\affiliation{Department of Electrical \& Computer Engineering, University of Minnesota, Minneapolis, Minnesota, 55455, USA}
\author{Michael Sammon}
\affiliation{Department of Electrical \& Computer Engineering, University of Minnesota, Minneapolis, Minnesota, 55455, USA}
\author{Eugene Mele}
\affiliation{Department of Physics and Astronomy, University of Pennsylvania, Philadelphia, Pennsylvania, 19104, USA}
\author{Tony Low}
\email{tlow@umn.edu}
\affiliation{Department of Electrical \& Computer Engineering, University of Minnesota, Minneapolis, Minnesota, 55455, USA}

\date{\today}

\begin{abstract}
	Artificial lattices have been used as a platform to extend the application of topological physics beyond electronic systems. Here, using the two-dimensional Lieb lattice as a prototypical example, we show that an array of disks which each support localized plasmon modes give rise to an analog of the quantum spin Hall state enforced by a synthetic time reversal symmetry. We find that an effective next-nearest-neighbor coupling mechanism intrinsic to the plasmonic disk array introduces a nontrivial $Z_2$ topological order and gaps out the Bloch spectrum. A faithful mapping of the plasmonic system onto a tight-binding model is developed and shown to capture its essential topological signatures. Full wave numerical simulations of graphene disks arranged in a Lieb lattice confirm the existence of propagating helical boundary modes in the nontrivial band gap. 
	\end{abstract}

\maketitle

\section{Introduction}

Artificial lattices can be patterned into ordered structures designed to control efficient flow of energy and information\cite{Joannapoulos2011}.  There has been particular interest in developing periodic structures for this purpose that use nontrivial topology in their interior Bloch bands\cite{Lu2014a, Khanikaev2017}.  These are known to support protected transport channels at boundaries between distinct topological states  in analogy to the well-studied electronic surface states that occur at the boundaries of two and three dimensional topological insulators\cite{Hasan2010}. Indeed, nontrivial bulk Chern bands in two dimensions have been successfully developed on photonic\cite{Wang2009a}, plasmonic\cite{Jin2017c} and even mechanical platforms\cite{Nash2015}, where time reversal symmetry is broken in order to establish their topological nature. In striking contrast,  photonic 2D analogs to quantum spin Hall states which retain time reversal symmetry have been elusive. A key challenge is that the Kramers degeneracies for half-integral angular momentum, whose connectivity is essential for defining the topological state, are generically absent from photonic and plasmonic analogs\cite{Lu2016}. While multiple realizations of a photonic quantum spin Hall state have been demonstrated\cite{Hafezi2011, Khanikaev2012, Wu2015}, intricate engineering via fine tuning of the design is required to enforce the required symmetries\cite{Khanikaev2017}.

In this work we suggest a robust resolution to this dilemma, and illustrate it with studies on a two dimensional plasmonic Lieb lattice\cite{Weeks2010, Goldman2011} as a prototype.  Our model is characterized by two length (energy) scales that express the confinement of excitations within and transmission between nodes of a Lieb lattice network. Over a realizable parameter range in which these scales are of comparable size, we identify both an emergent two-fold Kramers degeneracies and a synthetic spin-orbit process that gaps the Bloch spectra to endow the bands with nontrivial $Z_2$ topological order\cite{Kane2005a}. We test this idea with full wave numerical calculations to  confirm the existence of propagating helical modes on the boundaries and we develop a mapping onto a tight binding model that captures the essential spectral signature of this state.  We emphasize that these phenomena in artificial lattices are controlled by two energy scales which are tunable over a wide dynamic range and access ordered states that are practically unachievable in ordinary electronic materials. This strategy can be generalized to a wide family of appropriately designed artificial lattices.

\section{Results and discussion}

% Plasmonic band structure
A plasmonic band structure can be calculated when the graphene nanodisks are arranged in a periodic configuration. Here we consider placing graphene nanodisks on a Lieb lattice\cite{Goldman2011}. When separation between the disks, $a$, is much larger than the nanodisk diameter, $d$, coupling between the plasmon modes is negligible and we observe a ladder of on-site resonant multipole electromagnetic modes (see \cref{fig:1}a). The three groups of energy bands each correspond to the dipole, quadrupole, and hexapole modes of the graphene nanodisk. As the disk separation is decreased, coupling strength between plasmon modes increases and the dispersion in the plasmon band structure becomes stronger. In \cref{fig:1}b the plasmon band structure for $a/d=1.05$ is shown. Interestingly, we find that the dipole, quadrupole, and hexapole modes develop a non-trivial band structure resembling the band structure of a tight binding Lieb lattice. Finally, when the disks begin to overlap ($a<d$, \cref{fig:1}c), a stark change in the band structure is observed. A low energy hydrodynamic plasmon mode for which the dispersion satisfies $\omega \rightarrow 0 $ as $q\rightarrow 0$ appears and the mapping onto a tight-binding Lieb lattice breaks down. This situation corresponds to the setup given in previous works \cite{Silveiro2013, Xiong2019, Jin2017c, Jung2018} where periodic modulation of structure or density are used to perturb the hydrodynamic mode.

% Setting up the tight-binding model
To better understand the plasmon band structure given in \cref{fig:1}b, we develop a mapping of the plasmonic crystal onto a tight-binding model. The governing equations of the plasmonic response can be written as an eigenvalue equation $\hat{H}\psi =\omega\psi$ where\cite{Jin2017c}

\begin{equation}
	\hat{H}=
	\begin{pmatrix}
		0 & \hat{V} \hat{\mathbf{p}}^T\sqrt{\omega_F (\mathbf{r})} \\ \frac{e^2}{\pi\hbar} \sqrt{\omega_F(\mathbf{r})}\hat{\mathbf{p}}  & 0 
	\end{pmatrix}
	,\quad 
	\psi = 
	\begin{pmatrix}
		\Phi \\ \mathbf{J}/\sqrt{\omega_F(\mathbf{r})}
	\end{pmatrix}.
\end{equation}
The eigenstate $\psi$ is a vector of the potential $\Phi$ and current density $\mathbf{J}$, $\mathbf{\hat{p}}$ is the in-plane momentum operator, $\hat{V}$ is the Coulomb operator, and $\omega_F=E_F/\hbar$ is the Fermi frequency. The localized plasmonic modes of the nanodisk are analogous to the atomic orbitals in a tight-binding theory. From an eigenmode calculation of the isolated graphene disk, we find that the electric potential is given by
\begin{equation}
	\Phi_{n,l}^p(\mathbf{r}) =
	\begin{cases}
	u_n(r)\cos(l\theta),  & p=+\\ 
	u_n(r)\sin(l\theta), & p=- 
	\end{cases}
\end{equation}	
where $u_n(r)$ gives the radial dependence of $\Phi$, $n$ is the number of nodes in $u_n(r)$, and $l$ is a positive integer. Here we focus on the six lowest energy modes which have $n=0$ and $l=1,2,3$. Henceforth the superscript $n$ will be omitted for notational simplicity. Note that rotational symmetry of the isolated graphene disk implies that $\psi_l^+$ and $\psi_{l}^-$ are degenerate in energy. Details on the Hamiltonian and eigenstates are given in the supplementary information.

% Hopping parameters
Equipped with the Hamiltonian and atomic orbitals, we may now calculate the matrix elements of the tight-binding Hamiltonian. In general, the matrix elements between orbitals $\psi_l^p$ and $\psi_{l'}^{p'}$ on disks separated by $\mathbf{R}$ are given by
\begin{equation}\label{eq:tbparameters}
	t_{ll'}^{pp'}(\mathbf{R})= 
	 -i\frac{e^2}{\pi\hbar} \int 
	\Bigl(
	\mathbf{J}^{p*}_l (\mathbf{r}) \cdot \nabla\Phi_{l'}^{p'}(\mathbf{r-R})+\Phi^{p*}_l(\mathbf{r})\nabla\cdot \mathbf{J}_{l'}^{p'}(\mathbf{r-R})
	\Bigr)d^2r
\end{equation}
where the inner product is defined such that $\hat{H}\psi=\omega\psi$ is a Hermitian eigenproblem\cite{Jin2017c} (see supplementary information for details). Given that we have the current density and electric potential of the orbital modes, it is straightforward to calculate the tight-binding model parameters. For the Lieb lattice, the nearest neighbor (NN) hopping parameters can be calculated by setting $\mathbf{R}=\pm a\mathbf{\hat{x}}\equiv \pm \mathbf{a}_x$ or $\mathbf{R}=\pm a\mathbf{\hat{y}}\equiv \pm  \mathbf{a}_y$ in \cref{eq:tbparameters}.

% Quadrupole NN tight-binding Hamiltonian
We now take a closer look at the nearest neighbor coupling between the quadrupole orbitals $\psi_{2}^{\pm}$. Either by explicit calculation through \cref{eq:tbparameters} or by inspection of the mode symmetries, it may be shown that hopping from $\psi_{2}^+$ to $\psi_{2}^-$ given by $t_{22}^{+-}(\mathbf{a}_j)$ is zero for $j=x,y$. It follows from the $C_4$ symmetry of the quadrupole modes that hopping amplitudes in the $x$ and $y$ directions will be equal, i.e. $t_{22}^{pp}(\mathbf{a}_x)=t_{22}^{pp}(\mathbf{a}_y)$ for $p=\pm$. Finally, it is important to note that the $\psi_{2}^+$ and $\psi_{2}^-$ orbitals have hopping amplitudes with opposite sign. The plasmonic band structure of the quadrupole with nearest neighbor hopping will therefore have two copies of the the 3-band Lieb lattice band structure, where one of the copies will be inverted in energy with respect to the other. This is indeed what we observe for the quadrupole modes as shown in \cref{fig:1}b. A similar analysis can be applied to the other multipole modes. Here we will be focusing on the quadrupole mode band structure since it shares the $C_4$ symmetry of the lattice. 

\begin{figure}
	\centering
	\includegraphics{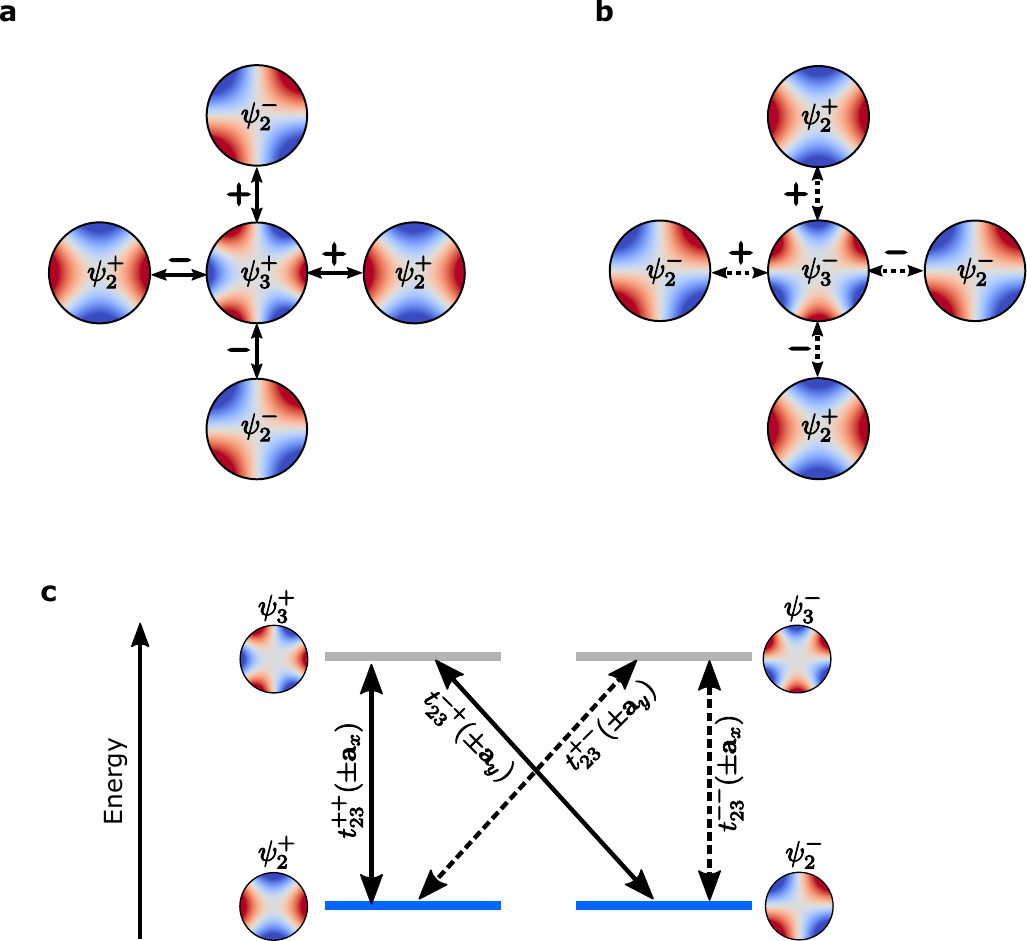}
	\caption{\textbf{Orbital coupling between $\psi_{2}^\pm$ and $\psi_{3}^\pm$}. \textbf{a}, Representation of nearest neighbor hopping between $\psi_{3}^+$ and $\psi_{2}^\pm$. Hoppings for $\mathbf{R}=\pm \mathbf{a}_x,\pm\mathbf{a}_y$ are considered. The electric potential $\Phi$ of the modes are shown. The plus and minus signs associated with each hopping parameter indicate its sign. \textbf{b}, Hopping between $\psi_{3}^-$ and $\psi_{2}^\pm$ are shown. All hopping parameters for $\textbf{b}$ can be generated by rotation of the configuration in \textbf{a}. \textbf{c}, The allowed hopping transitions are represented on an energy level diagram. A second order transition connecting $\psi_{2}^+$ to $\psi_{2}^-$ is possible via the $\psi_{3}^\pm$ modes.}
	\label{fig:2}
\end{figure}

% Hopping between l=2 and l=3
We now extend the model to include orbital coupling between $\psi_{3}^+$ and $\psi_{2}^\pm$.  Note that the same analysis can be applied to coupling between $\psi_{2}^\pm$ and $\psi_{1}^\pm$. In \cref{fig:2}a all allowed hoppings between the $\psi_{3}^+$ and $\psi_{2}^\pm $ modes are shown along with their respective signs. Hopping parameters $t_{23}^{++}(\pm \mathbf{a}_y)$, $t_{23}^{-+}(\pm \mathbf{a}_x)$ are zero and are not shown. Applying a $C_4$ rotation centered on the $\psi_{+3}$ mode transforms the configuration of \cref{fig:2}a into \cref{fig:2}b with an overall sign change on all sites. This transformation directly gives the hopping parameters between $\psi_{3}^-$ and $\psi_{2}^\pm$. All the allowed transitions are illustrated on an energy level diagram in \cref{fig:2}c. 

Considering second order hopping processes mediated by the $\psi_{3}^\pm$ orbital states, we find two processes that affect the quadrupole mode band structure. First, a second order hopping along two links in the $x$ or $y$ directions (e.g. $t_{23}^{++}(\mathbf{a}_x)$ followed by $t_{32}^{++}(\mathbf{a}_x)$) provides a uniform shift in the energy of all orbital modes. A more interesting effect is found when a left or right turn is made between the first and second hops. Taking into account the signs given in \cref{fig:2}, we arrive at an effective second neighbor hopping given by
\begin{equation}\label{eq:tb_so}
	W = it_s\sum_{pp'}\sum_{\langle\langle ij\rangle\rangle}v_{ij}c^\dagger_{ip}\left[\sigma_y\right]_{pp'} c_{jp'}
\end{equation}
where $v_{ij}=(\mathbf{d}_1 \times \mathbf{d}_2)_z$, $\mathbf{d}_{1,2}$ are the nearest neighbor bonds connecting site $i$ to $j$, $c^\dagger _{ip}$ is a creation operator for orbital $\psi_2^p$ on lattice site $i$, and $\sigma_y$ is a Pauli matrix acting on the quadrupole orbital degrees of freedom. This intrinsic interaction describing a coupling between the $\psi_{2}^\pm$ orbitals that is path-dependent is similar to the Kane-Mele spin-orbit interaction discussed in the context of electronic systems\cite{Kane2005a}. Strength of the interaction $t_s$ can be found by projecting the full Hamiltonian into an effective Hamiltonian in the $\psi_{2}^\pm$ basis. The effective Bloch Hamiltonian for the $\psi_{2}^\pm$ orbital basis may then be written as
\begin{equation}\label{eq:tb_bloch}
	h(\mathbf{k})=
	\begin{pmatrix}
		h_{+}(\mathbf{k}) & w(\mathbf{k}) \\ w^\dagger(\mathbf{k}) & h_{-}(\mathbf{k})
	\end{pmatrix}=
	\begin{pmatrix}
		h_{+}(\mathbf{k}) & w(\mathbf{k}) \\ w^\dagger(\mathbf{k}) & \alpha h_{+}(\mathbf{k})
	\end{pmatrix}
\end{equation}
where $h_\pm$ is the intra-orbital Hamiltonian for orbital $\psi_{2}^\pm$, $\alpha$ is the ratio between hopping amplitudes $t_{22}^{++}$ and $t_{22}^{--}$, and $w$ is a skew-Hermitian Hamiltonian that describes the effective second neighbor coupling given by \cref{eq:tb_so}. Matrix representations of \cref{eq:tb_bloch} are given in the supplementary information. With the effective Bloch Hamiltonian, we are able to map the quadrupole degrees of freedom of a plasmonic lattice onto a simple tight-binding model with two basis orbitals on each lattice site (see \cref{fig:3}a,b).

% Topological classification for \alpha = -1 case
For the ideal case with $\alpha=-1$ in \cref{eq:tb_bloch}, the Hamiltonian may be written as
\begin{equation}\label{eq:kramers_h}
	h(\mathbf{k})=h_+(\mathbf{k})\otimes \sigma_z + w(\mathbf{k})\otimes i\sigma _y.
\end{equation} 
The band structure for this Hamiltonian is shown in \cref{fig:3}c. A strict topological classification of this Hamiltonian based on its symmetries is possible. Define a synthetic time-reversal operator as $T=\lambda_z\otimes i \sigma_y K$ where $\lambda_z=\textrm{diag}(1,-1,1)$ is a sublattice symmetry operator acting on the sublattice degrees of freedom, $\sigma_y$ is a Pauli matrix acting on the orbital degrees of freedom, and $K$ represents complex conjugation. Then, the Hamiltonian in \cref{eq:kramers_h} satisfies $Th(\mathbf{k})T^{-1}=h(-\mathbf{k})$. Since $T^2=-1$, a two-fold degeneracy is enforced at the time reversal invariant momenta by Kramer's theorem. A particle-hole symmetry operator can also be defined as $C=\lambda_z\otimes \sigma_z K$ such that $Ch(\mathbf{k})C^{-1}=-h(-\mathbf{k})$ and $C^2=+1$. From the time-reversal and particle-hole symmetry, we can now place the $\alpha=-1$ Hamiltonian into the DIII symmetry class\cite{Chiu2016}. In two-dimensions, this class is characterized by a $Z_2$ topological invariant. Since inversion symmetry is preserved, we may calculate the $Z_2$ invariant\cite{Fu2007a} $\nu$ of a single band by
\begin{equation}
(-1)^\nu = \prod_i \delta_i, \quad \delta_i = \xi(\Gamma _i)
\end{equation}
where $\Gamma_i$ are the time reversal invariant momenta in the Brillouin zone and $\xi(\Gamma_i)$ is the parity eigenvalue of the band at $\Gamma_i$. Unlike the electronic case for which $\delta_i$ is given as a product of the parity eigenvalues for all occupied bands, here we associate a separate $\nu$ with each band individually since there is no notion of band filling. The $Z_2$ invariant for each band shown in \cref{fig:3}c reveals the topologically non-trivial nature of the ideal plasmonic Lieb lattice.

% Comparison with physical case (\alpha=-1.2)
Physically, $\alpha = -1$ requires fine tuning to a special state where the symmetry is strictly enforced. The plasmonic lattice comprised of disks does not have symmetries that enforce the $\alpha=-1$ condition. When $\alpha\neq -1$, classification of the system into the DIII class no longer holds and we do not have a sharp definition for the $Z_2$ topological invariant. In addition, topologically trivial intra-orbital second neighbor coupling is also present in the plasmonic lattice. Tight-binding calculations for the bulk band structure with $\alpha=-1.2$ and the trivial second neighbor coupling are shown in \cref{fig:3}d. Breaking of the synthetic time-reversal symmetry is evident from the lifted degeneracy of the bulk bands at TRIM points $\Gamma$ and $X$. The spectrum is also no longer symmetric with respect to $E=0$ because particle-hole symmetry is broken by the trivial second neighbor hopping. 

However, by examining the edge states of the physical $\alpha\neq -1$, we find that the topological signatures of the ideal lattice are still present. The spectrum of the edge modes can be studied in a finite system ribbon geometry and shows that the edge modes from the ideal DIII structure (\cref{fig:4}a) survive even when the strict symmetry classification breaks down (\cref{fig:4}b). A difference between the two cases is observed at the edge mode crossing points shown in \cref{fig:4}d,e where the symmetry broken case shows an avoided crossing originating from mixing of the edge modes. In principle, propagating edge modes near this crossing point will be able to back-scatter. However, the size of the edge mode avoided crossing is roughly two orders of magnitude smaller than the band width, making it negligible in most practical situations. Full wave numerical simulations of the bulk plasmonic band structure shown in \cref{fig:3}e confirm that while the $\alpha=-1$ condition is not strictly satisfied, the edge modes of topological origin still survive as shown in \cref{fig:4}c,f. 

% Basis transformation and edge states
To elucidate the properties of the edge modes, we apply a transformation into the angular momentum basis to \cref{eq:tb_bloch}. The angular momentum basis states are given by $\tilde{\psi}_m=\psi_{|m|}^++i\textrm{sgn}(m)\psi_{|m|}^-$ where the electric potential of $\tilde{\psi}_m$ is $\tilde{\Phi}_m(\mathbf{r})\equiv u_0(r)e^{im\theta}$ and $m=\pm 1, \pm 2, \pm 3$. Applying the transformation to \cref{eq:tb_bloch} gives the tight binding Hamiltonian in the angular momentum basis
\begin{equation}\label{eq:tb_lbasis}
	\tilde{h}(\mathbf{k})=
	\begin{pmatrix}
		\bar{h}(\mathbf{k})+iw(\mathbf{k}) & \Delta(\mathbf{k}) \\ \Delta(\mathbf{k}) & \bar{h}(\mathbf{k})-iw(\mathbf{k})
	\end{pmatrix}
\end{equation}
where $\bar{h}=(h_{+}+h_{-})/2$ and $\Delta = (h_{+}-h_{-})/2$. Written in this form, it is clear that the plasmonic crystal will behave in a similar manner to the electronic quantum spin-hall insulator where the electronic spin degrees of freedom is mapped onto the chirality of the plasmon modes. The plasmonic edge modes observed in \cref{fig:4}f will have a potential proportional to $e^{\pm i2\theta}$ on each disk where the chirality is tied to the propagation direction.

% Edge state excitation in plasmon lattice
The existence of helical edge states can be confirmed by a simulation on a finite array of graphene disks. In order to couple with the $\tilde{\psi}_{-2}$ angular momentum mode, we use a circularly polarized dipole with dipole moment $\mathbf{p}=\hat{\mathbf{x}}+i\hat{\mathbf{y}}$ placed at the edge of an A site graphene disk. Simulation results in \cref{fig:5} show a projection of the excited plasmon fields onto the $\tilde{\psi}_{\pm 2}$ modes. We find that the dipole strongly couples to a $\tilde{\psi}_{-2}$ mode which propagates counter-clockwise along the boundaries of the system. A weak excitation of $\tilde{\psi}_{+2}$ is caused by imperfect coupling between the circularly polarized dipole and the $\tilde{\psi}_{-2}$ mode. 

% Trivial perturbations
Finally, we identify two other perturbations that open a topologically trivial gap at the M point and may occur in an array of plasmonic disks. The first is an inversion breaking perturbation which may result from an asymmetric hopping along either the $x$ or $y$ links in the lattice. A shift of the A or C site plasmonic disk towards the B site will give rise to the inversion breaking perturbation. Shifting the B site energy also opens up a gap on one side of the flat bands. A shift in the Fermi energy of the plasmonic disk will result in this type of perturbation. When applying either of the above mentioned perturbations to an ideal DIII class system, the edge modes are removed only after the gap closes and reopens. Interestingly, even in the physical case where we are no longer in the ideal DIII class, the edge modes persist until the gap is closed, implying that the edge modes remain robust to external perturbations. Results on perturbations are given in the supplementary information.

\section{Conclusion/Discussion}
% Summary
In summary, we have shown that propagating helical edge modes induced by an intrinsic synthetic spin-orbit process exists in an artificial plasmonic crystal. We consider the Lieb lattice as a prototypical example and develop a mapping of the plasmonic system onto a simple tight-binding model. With the tight-binding model, we identify an ideal limit of the system in which it can be classified into the DIII symmetry class. In this limit, a non-trivial $Z_2$ invariant is calculated and the associated topologically protected edge states are shown to exist. It is important to note that this ideal limit is distinct from the case of electron spins on a Lieb lattice and is a result of the opposite sign hopping between the localized plasmon modes. Although the physical plasmonic lattice does not strictly follow the behavior of this ideal limit, we are able to verify that the system is close enough to retain the edge states found for the ideal DIII limit. Propagation of the helical plasmonic edge modes is verified using full wave numerical simulations.

% Considerations for experimental realization
Experimentally, the proposed helical edge modes can be excited by placing a near-field antenna at the edge of the plasmon disk array as shown in \cref{fig:5}c. For the setup used in this work, the gap in which the edge mode can be observed is 4meV. We expect the edge mode in this gap to be observable as a plasmonic band gap of similar size has been successfully resolved using the near-field scanning microscope technique\cite{Xiong2019}. The energy scale of the plasmonic band structure can also be shifted by simply scaling the geometry of the lattice (see supplementary information). For a disk size of $d=700$nm and periodicity $a=735$nm, the quadrupole band energies are lowered to 70meV. Importantly, the ratio between bandwidth and band gap stays constant as the geometry is scaled. Hence the physical phenomena that we have examined for the setup in this work can be translated to a different energy range by scaling the geometry accordingly. 

\section*{Methods}
\begin{figure}
	\centering
	\includegraphics{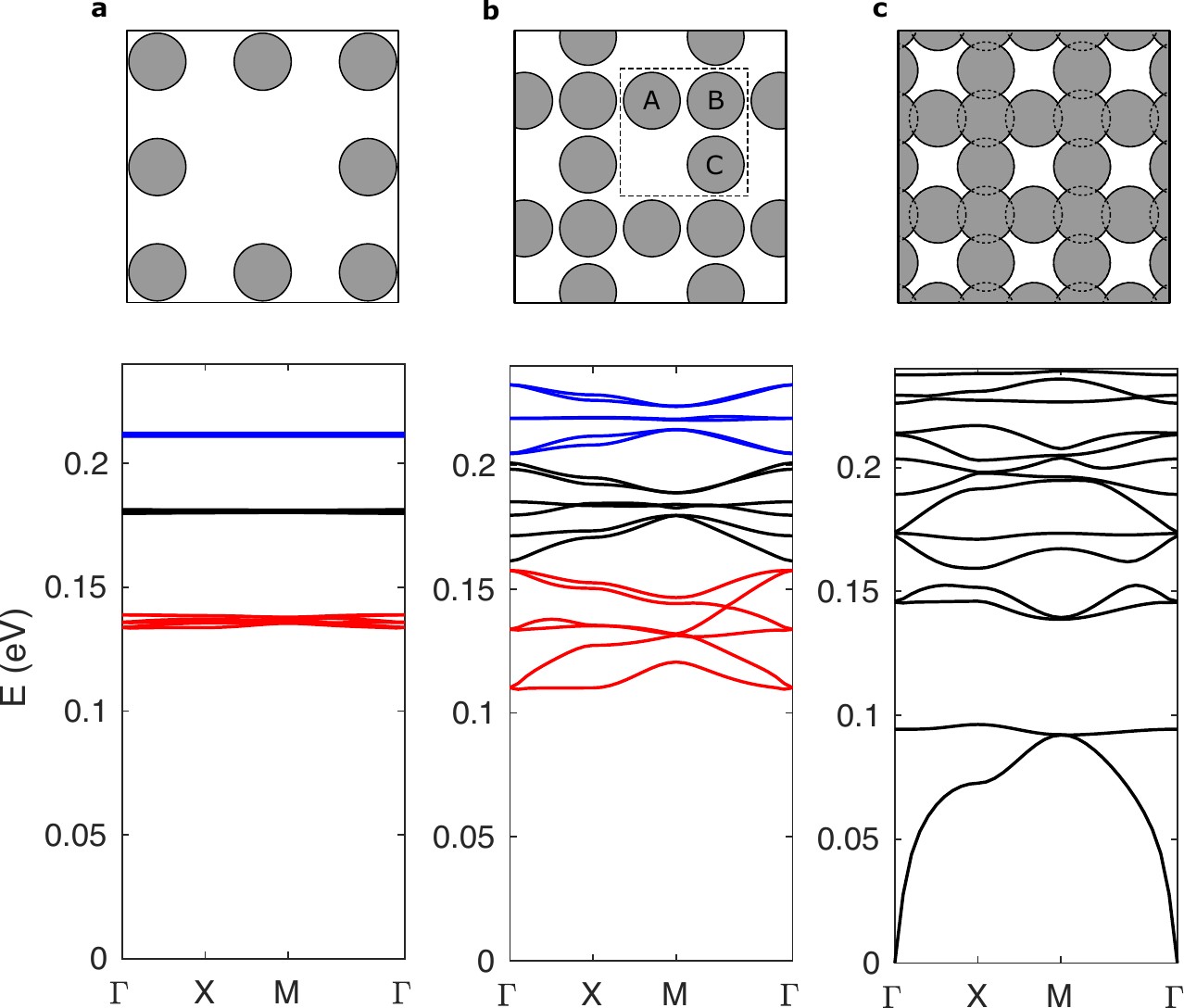}
	\caption{\textbf{Plasmonic band structure of graphene nanodisk arrays.} \textbf{a}, Array schematic and plasmonic band structure of graphene nanodisk array with disk separation $a=200$nm and disk radius $d=100$nm. Bands corresponding to the dipole(red), quadrupole(black), and hexapole(blue) are shown in the band diagram. \textbf{b}, The disk separation is reduced to $a=105$nm while disk radius is fixed to $d=100$nm. The dashed box indicates the unit cell. Bands formed from the dipole, quadrupole, and hexapole modes of the graphene nanodisk with the same color coding as \textbf{a}. \textbf{c}, For $a=85$nm and $d=100$nm the disks are now overlapping. Color coding of the bands is not applied since classification in terms of the multipole modes of the nanodisk does not hold in this case. For all simulations, the graphene Fermi energy is set to $E_F=0.5$eV and the substrate dielectric constant is $\epsilon_{s}=2.2$.}
	\label{fig:1}
\end{figure}

\begin{figure}
	\centering
	\includegraphics[width=\textwidth]{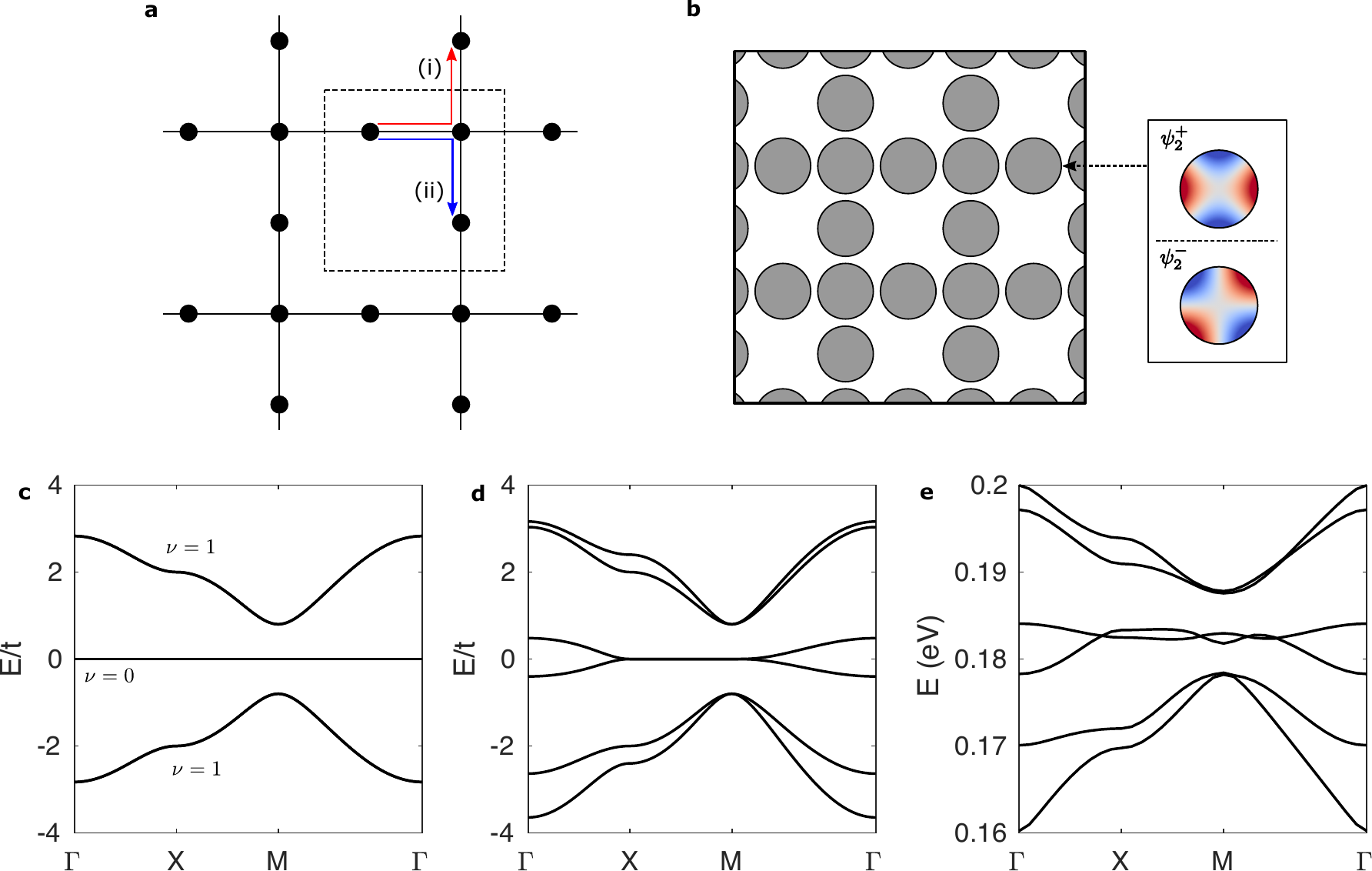}
	\caption{\textbf{Tight binding mapping of a plasmonic Lieb lattice.} \textbf{a}, Tight binding model onto which the plasmonic crystal can be mapped onto. The red and blue arrows represent two paths through which the interaction \cref{eq:tb_so} connects a $\psi_{2}^+$ state to a $\psi_{2}^-$ state.  \textbf{b}, Schematic representation of a graphene nanodisk Lieb lattice with $\psi_{2}^+$ and $\psi_{2}^-$ orbitals on each site. Parameters are equivalent to the setup in \cref{fig:1}b. \textbf{c,d}, Band structure calculated from tight binding model given in \cref{eq:tb_bloch} with $\alpha=-1$ and $\alpha=-1.2$ respectively. For the $\alpha=-1$ case, the $Z_2$ topological invariant $\nu$ for each band is also shown. For the $\alpha=-1.2$ case, trivial second neighbor hopping is included. \textbf{e}, Full wave electromagnetic simulation of plasmonic band structure for configuration shown in \textbf{b}. }
	\label{fig:3}
\end{figure}

\begin{figure}
	\centering
	\includegraphics[width=\textwidth]{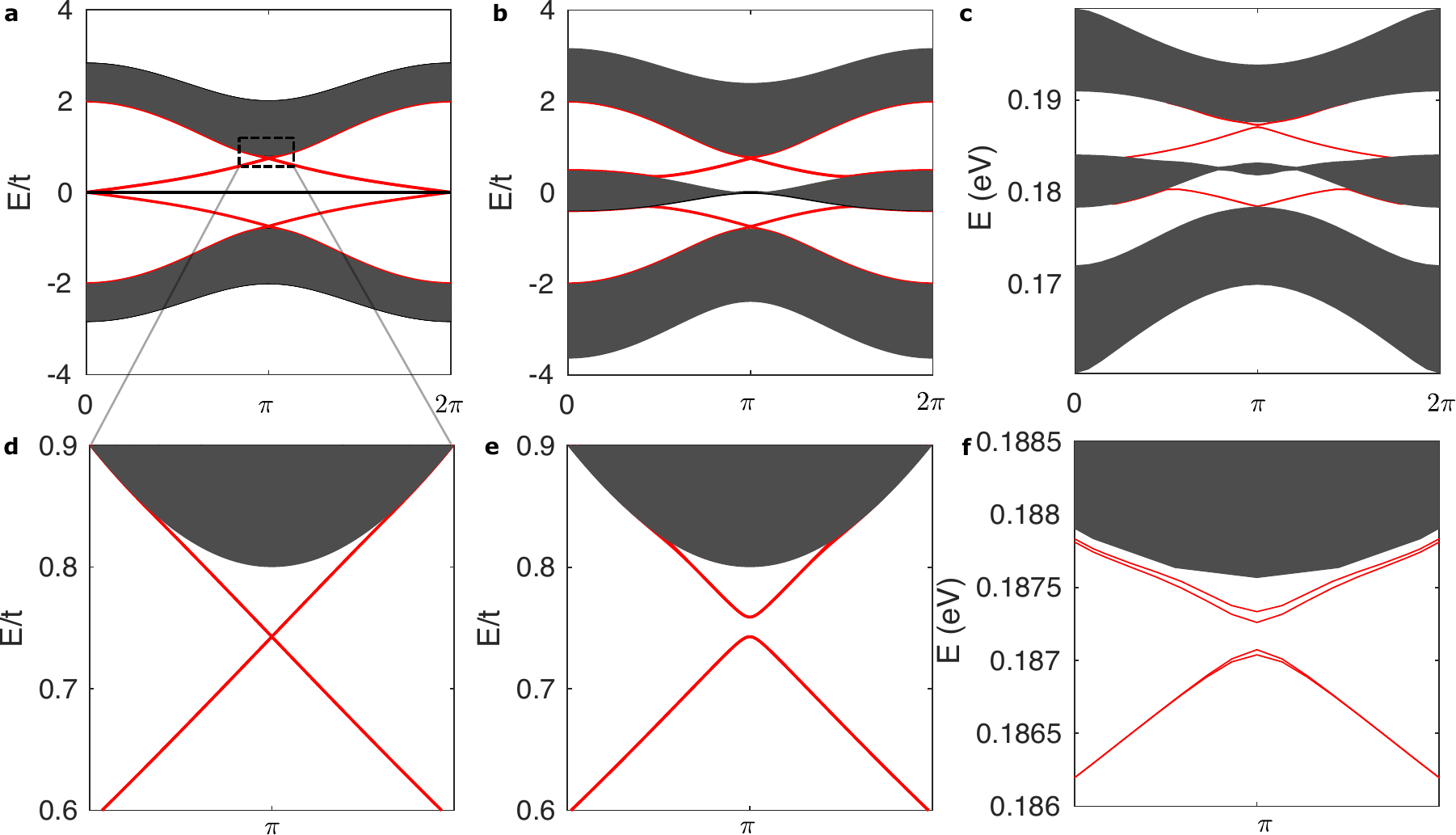}
	\caption{\textbf{Edge modes of the plasmonic Lieb lattice}. \textbf{a,d} Tight-binding band structure calculation for Lieb lattice in a ribbon geometry with $\alpha=-1$. The edge modes are highlighted in red while the bulk band projections are colored in gray. All edge modes are doubly degenerate. A more detailed view of the crossing point of \textbf{a} is given in \textbf{d}. \textbf{b,e} Tight-binding band structure for $\alpha=-1.2$ and with trivial second neighbor hopping included. \textbf{c,f} Full wave numerical simulation for plasmonic Lieb lattice in ribbon geometry. Parameters are identical to those used for \cref{fig:1}b and \cref{fig:3}e}
	\label{fig:4}
\end{figure}

\begin{figure}
	\centering
	\includegraphics[width=\textwidth]{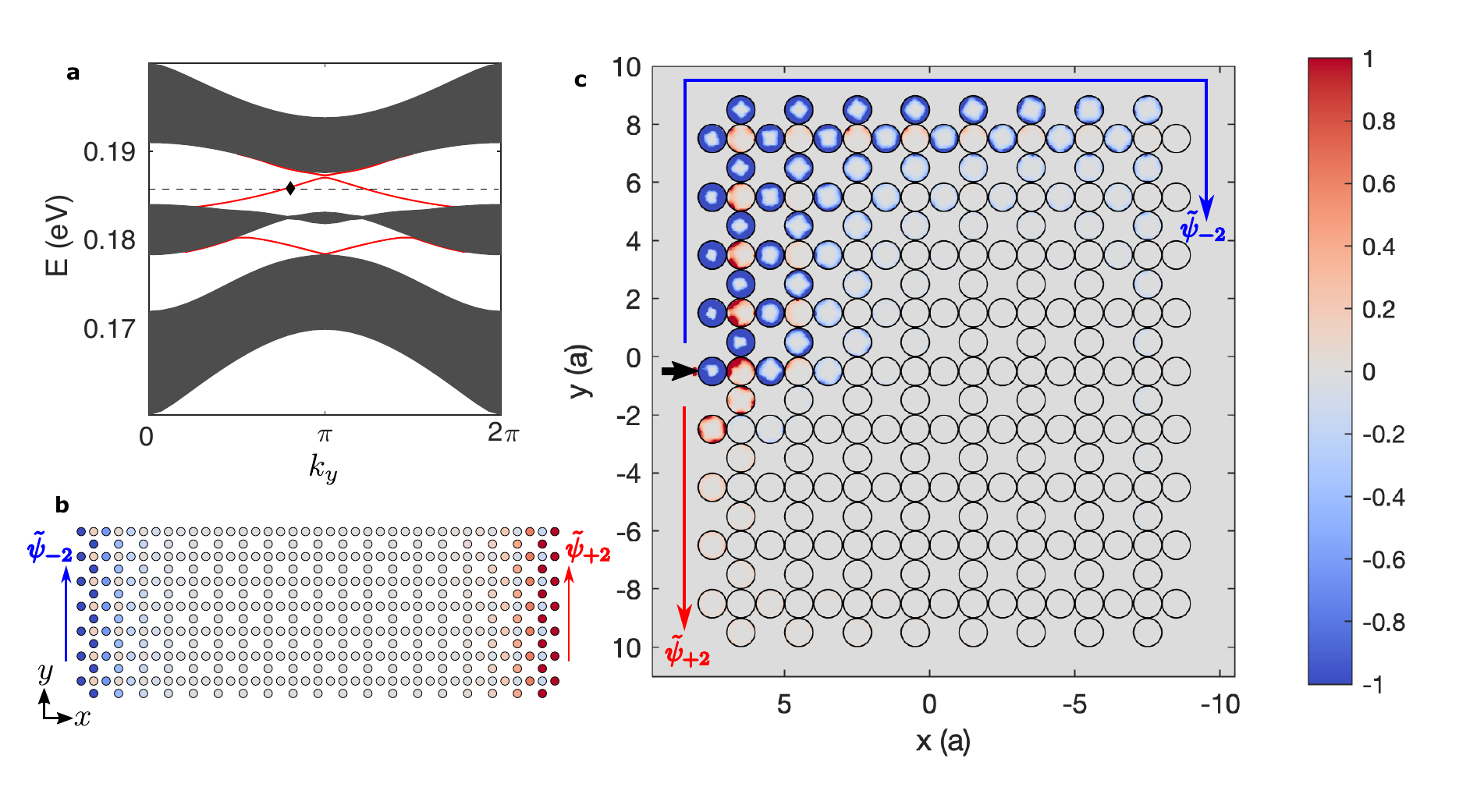}
	\caption{\textbf{Dipole launching the edge states of the plasmonic Lieb lattice.}\textbf{a}, Band structure of the plasmonic Lieb lattice in a ribbon geometry. Bands traversing the bulk gap are all doubly degenerate. \textbf{b}, Schematic representation of the eigenstate corresponding to the black diamond in \textbf{a}. The blue(red) shows projection onto state $\tilde{\psi}_{-2}$($\tilde{\psi}_{+2}$). Arrows indicate the group velocity of each edge mode. \textbf{c}, Simulation of topological edge plasmon launched by circularly polarized dipole at frequency indicated in \textbf{a}. Magnitude of the plasmon electric field is multiplied by a projection onto the $\tilde{\psi}_{\pm 2}$ angular momentum states. Dipole source is placed at black arrow 5nm above the graphene nanodisk. }
	\label{fig:5}
\end{figure}

All electromagnetic simulations were performed using COMSOL Multiphysics. Graphene is modeled as a surface current with a Drude conductivity, $\sigma(\omega)=\frac{e^2}{\pi\hbar}\frac{E_F}{\hbar\gamma-i\hbar\omega}$. Scattering time is set to $\tau=1$ps where $\gamma=1/\tau$. The bulk plasmonic band structure shown in \cref{fig:1} and \cref{fig:3} was performed by assuming periodic boundary conditions along both the in-plane directions. In the out-of-plane direction, the structure is padded with 1$\mu$m of free space and a perfectly matched layer of thickness 300nm. The plasmonic band structure calculation for a ribbon geometry shown in \cref{fig:4}c,f was performed using a ribbon length of 25 unit cells with periodic boundary conditions only in the direction parallel to the ribbon edges. The circularly polarized dipole used for \cref{fig:5}c was placed 5nm above the edge of the graphene nanodisk. Projection of the potential onto the angular momentum states $\tilde{\psi}_{\pm 2}$ was calculated as $\int e^{i2\theta}\Phi(\mathbf{r})/|\Phi(\mathbf{r})|d^2r - \int e^{-i2\theta}\Phi(\mathbf{r})/|\Phi(\mathbf{r})|d^2r$ for each disk. 

\begin{acknowledgements}
	SHP, MS and TL acknowledge support by the National Science Foundation, NSF/EFRI Grant No. EFRI-1741660. Work by EJM was supported by the Department of Energy under grant DE-FG02-84ER45118. SHP acknowledges partial support by the National Science Foundation through the University of Minnesota MRSEC under Award Number DMR-2011401.
\end{acknowledgements}

% References

\clearpage
% Figures

\end{document}

\title{Helical boundary modes from synthetic spin in a plasmonic lattice: supplementary information}
\author{Sang Hyun Park}
\affiliation{Department of Electrical \& Computer Engineering, University of Minnesota, Minneapolis, Minnesota, 55455, USA}
\author{Michael Sammon}
\affiliation{Department of Electrical \& Computer Engineering, University of Minnesota, Minneapolis, Minnesota, 55455, USA}
\author{Eugene Mele}
\affiliation{Department of Physics and Astronomy, University of Pennsylvania, Philadelphia, Pennsylvania, 19104, USA}
\author{Tony Low}
\email{tlow@umn.edu}
\affiliation{Department of Electrical \& Computer Engineering, University of Minnesota, Minneapolis, Minnesota, 55455, USA}
%\date{\today}

\maketitle

\section{Tight-binding mapping}
Plasmons on a conducting sheet with conductivty $\sigma(\omega,\mathbf{r})$ are described by the equations
\begin{equation}
	\rho(\mathbf{r})=\frac{1}{i\omega}\nabla\cdot\mathbf{J}(\mathbf{r})=\frac{-1}{i\omega}\nabla\cdot[\sigma(\omega,\mathbf{r})\nabla\Phi(\mathbf{r})],\quad  \Phi=\Phi_{ext} + \int \frac{\rho(\mathbf{r})}{|\mathbf{r-r'}|}d^2r'.
\end{equation}
To map the plasmonic lattice onto a tight-binding model, we need to cast the governing equations given above into an eigenvalue problem. Assuming a Drude model for the conductivity $\sigma(\omega,\mathbf{r})=i\frac{e^2 E_F}{\pi\hbar^2\omega}=i\frac{e^2}{\pi\hbar}\frac{\omega_F}{\omega}$ and following the formulation given in \cite{Jin2017c}, the governing equations can be written as $\hat{H}\psi=\omega \psi$ with
\begin{equation}
\hat{H}=
\begin{pmatrix}
0 & \hat{V}\mathbf{\hat{p}}^T\sqrt{\omega_F(\mathbf{r})} \\ \frac{e^2}{\pi\hbar}\sqrt{\omega_F(\mathbf{r})}\mathbf{\hat{p}} & 0
\end{pmatrix},\ 
\psi=
\begin{pmatrix}
\Phi(\mathbf{r}) \\ \mathbf{J}(\mathbf{r})/\sqrt{\omega_F(\mathbf{r})}
\end{pmatrix}.
\end{equation} 
The Coulomb and momentum operators are defined as $\hat{V}[f](\mathbf{r})=\int f(\mathbf{r})/|\mathbf{r}-\mathbf{r}'|d^2r'$ and $\hat{\mathbf{p}}=-i\nabla$.
The Hamiltonian can be decomposed into $\hat{H}=\hat{B}^{-1}\hat{A}$ where
\begin{equation}
	\hat{B}^{-1}=
\begin{pmatrix}
\alpha^{-1} \hat{V} & 0 \\ 0 & 1
\end{pmatrix}, \quad
\hat{A}=\begin{pmatrix}
0 & \frac{e^2}{\pi\hbar}\hat{\mathbf{p}}^T\sqrt{\omega_F(\mathbf{r})} \\ \frac{e^2}{\pi\hbar}\sqrt{\omega_F(\mathbf{r})}\mathbf{\hat{p}} & 0
\end{pmatrix}.
\end{equation}
The eigenvalue equation may then be written as $\hat{A}\psi = \omega \hat{B}\psi$ which is a generalized Hermitian eigenproblem with real eigenvalues and eigenstates orthogonal under the inner product $\langle\psi_\nu|\psi_\mu\rangle \equiv \int \psi_\nu^\dagger (\mathbf{r})\hat{B}\psi_\mu(\mathbf{r})d^2r$. Using the inner product definition, an element of the tight-binding Hamiltonian is calculated as 
\begin{equation}
	\langle \psi_\nu|\hat{H}|\psi_\mu\rangle =\int \psi_\nu^\dagger (\mathbf{r})\hat{B}\hat{H}\psi_\mu(\mathbf{r})d^2\mathbf{r}=\int \psi_\nu^\dagger(\mathbf{r})\hat{A}\psi_\mu(\mathbf{r})d^2\mathbf{r}=-i\frac{e^2}{\pi\hbar}\int \left(\mathbf{J}^*_\nu \cdot \nabla\Phi_\mu+\Phi_\nu^*\nabla\cdot\mathbf{J}_\mu \right)d^2r
\end{equation}

\begin{figure}
	\centering
	\includegraphics[width=0.8\textwidth]{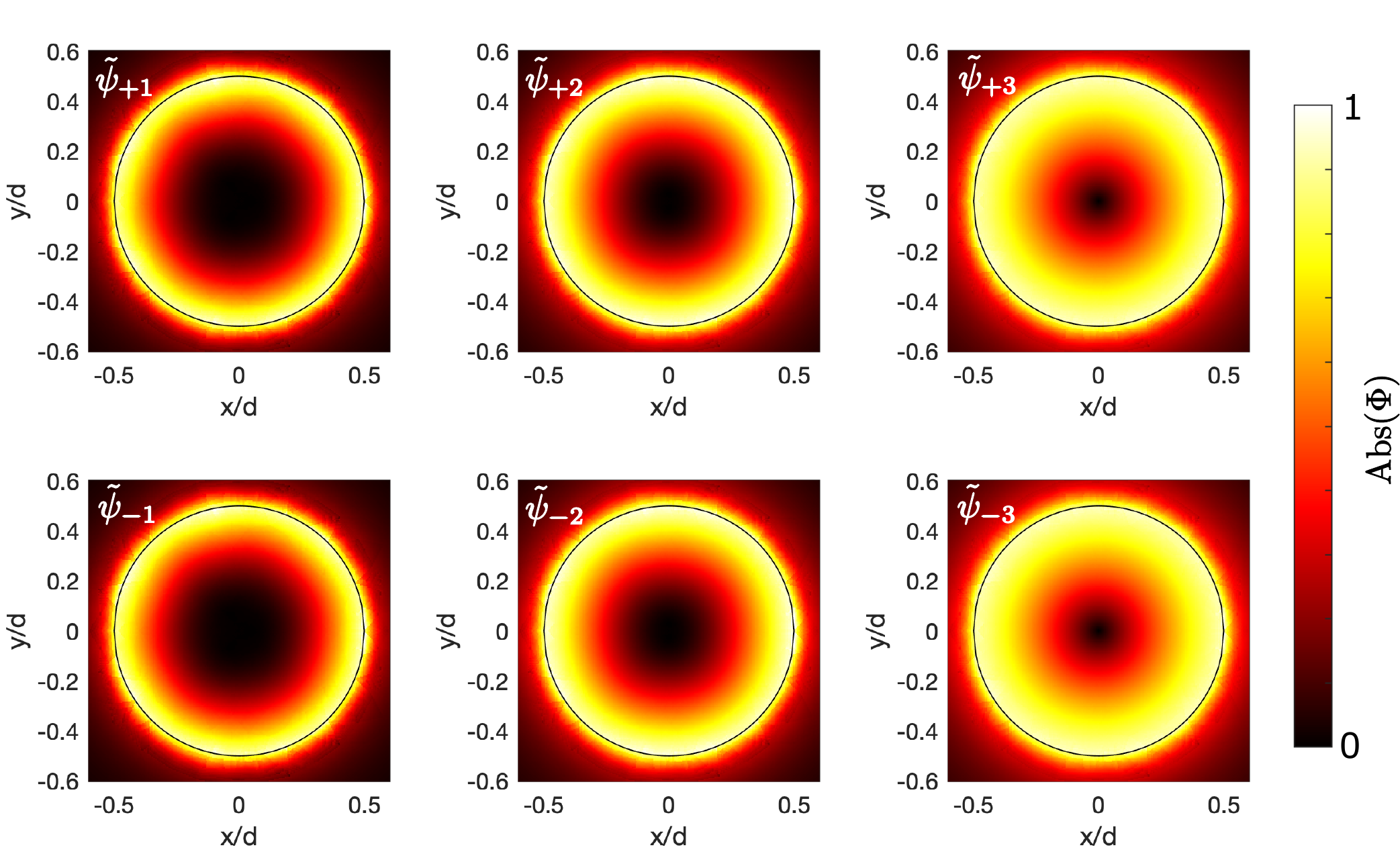}
	\caption{Magnitude of potential for $\tilde{\psi}_m$}
	\label{fig:Lbasis_mag}
\end{figure}

\begin{figure}
	\centering
	\includegraphics[width=0.8\textwidth]{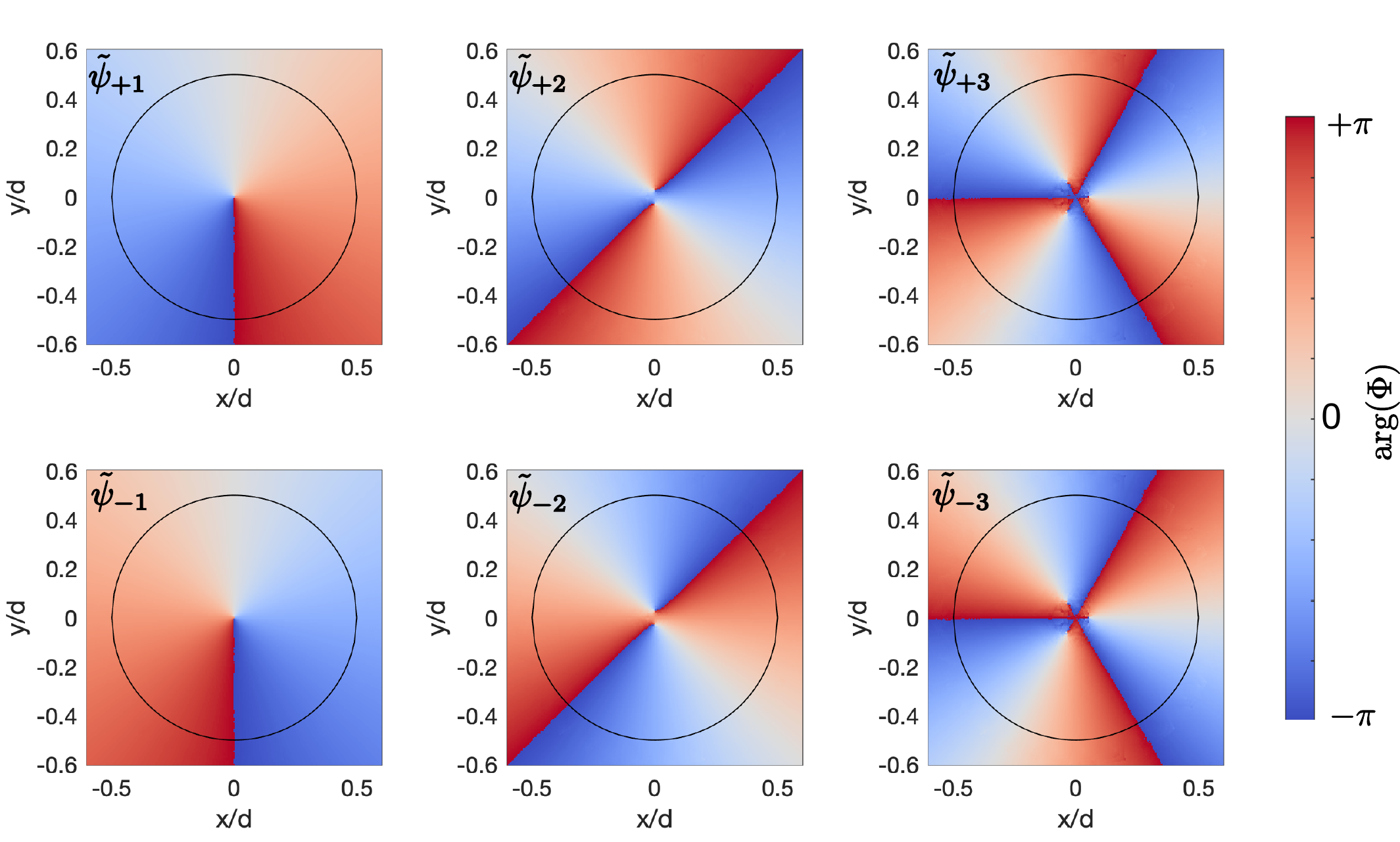}
	\caption{Phase of potential for $\tilde{\psi}_m$}
	\label{fig:Lbasis_phase}
\end{figure}

The eigenstates of an isolated plasmonic disk are calculated numerically using COMSOL. The potential of the $\psi_l^p$ eigenstates are shown in supplementary \cref{fig:realbasis}. Note that these eigenstates have a purely real potential. In the main text, we also define an angular momentum basis as $\tilde{\psi}_m=\psi_{|m|}^+ + i\textrm{sgn}(m)\psi_{|m|}^-$. The potential of the angular momentum basis states are complex. The magnitude and phase of the potential for the angular momentum eigenstates are shown in \cref{fig:Lbasis_mag} and \cref{fig:Lbasis_phase}.

\section{Matrix representation of Bloch Hamiltonian}
The matrix representation of the Bloch Hamiltonian (see eq.5 of main text) is given. The intra-orbital Hamiltonian with nearest-neighbor coupling is
\begin{equation}
	h_{\pm}(\mathbf{k})=t_{22}^{\pm\pm}
\begin{pmatrix}
0 & 1+e^{-ik_x} & 0 \\ 1+e^{ik_x}& 0 & 1+e^{-ik_y} \\ 0& 1+e^{ik_y}& 0
\end{pmatrix}.
\end{equation}
The trivial second-neighbor hopping adds a term proportional to 
\begin{equation}
\begin{pmatrix}
0 &0 & 1+e^{-ik_y}+e^{-ik_x}+e^{-ik_x-ik_y} \\ 0 & 0 & 0 \\ 1+e^{ik_y}+e^{ik_x}+e^{ik_x+ik_x} &0 & 0
\end{pmatrix}	
\end{equation}
to the matrices $h_\pm(\mathbf{k})$.

The topologically non-trival second neighbor hopping described by eq.4 of the main text is given by
\begin{equation}
	w(\mathbf{k})=t_s
\begin{pmatrix}
0 &0 & 1-e^{-ik_y}-e^{-ik_x}+e^{-ik_x-ik_y} \\ 0 & 0 & 0 \\ -1+e^{ik_y}+e^{ik_x}-e^{ik_x+ik_y} &0 & 0
\end{pmatrix}.
\end{equation}

\section{Eigenstates from numerical simulations}
\begin{figure}
	\centering
	\includegraphics[width=0.6\textwidth]{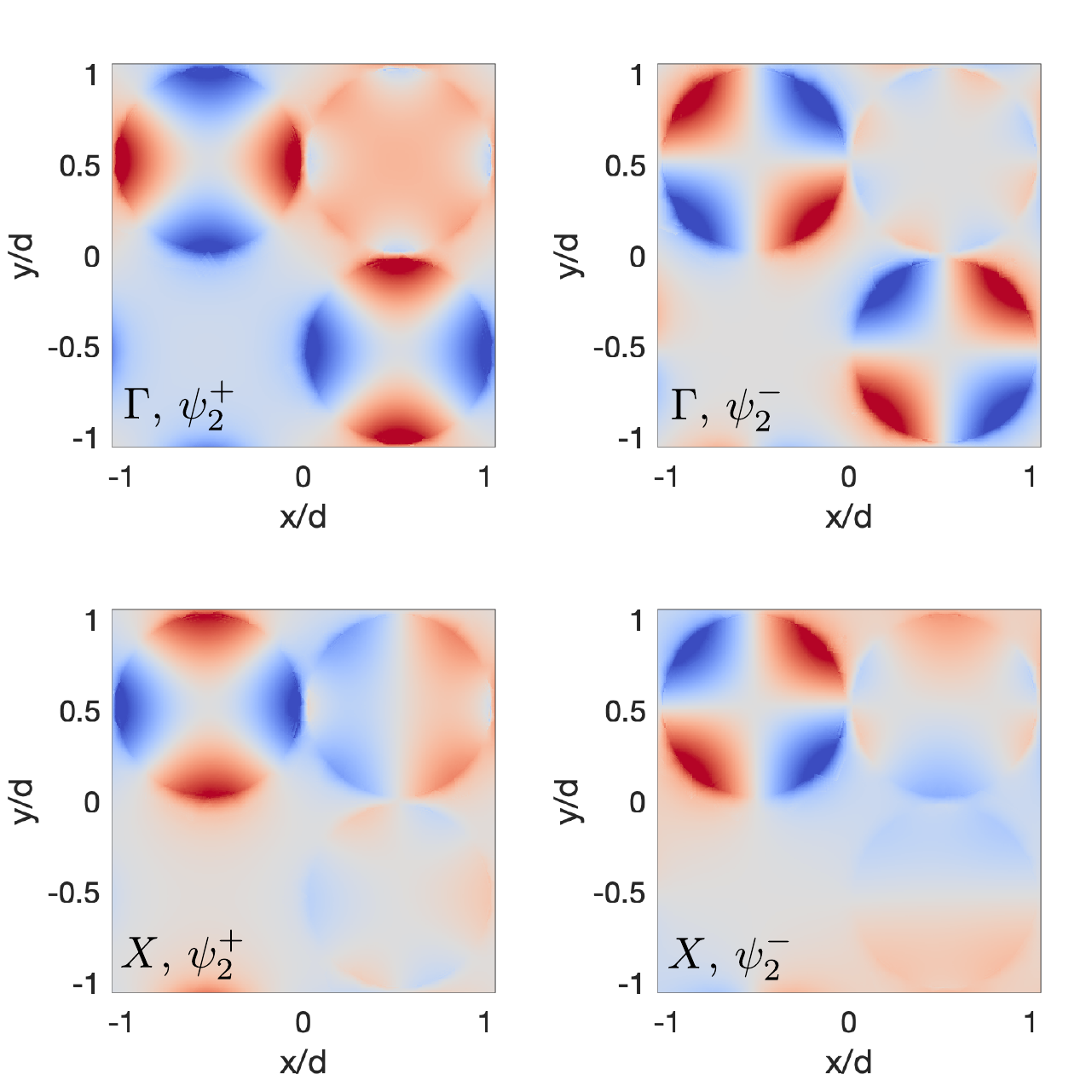}
	\caption{Eigenstates of the plasmon Lieb lattice}
	\label{fig:Lieb}
\end{figure}

The tight-binding mapping proposed in this work not only accurately describes the plasmonic band structure, but also correctly describes the eigenstates. For a Lieb lattice tight-binding model with only nearest neighbor hopping, the eigenstate of the flat band is given by $\psi^\dagger =t\left(\cos(k_y/2)a^\dagger_\mathbf{k}+\cos(k_x/2)c^\dagger_\mathbf{k}\right)$ where $a^\dagger_\mathbf{k},c^\dagger_\mathbf{k}$ are the creation operators for site A and C respectively. Therefore at the $\Gamma$ point we expect equal amplitude on the A and C sites while at the X point we will only have amplitude on the A site. Eigenstates of the plasmonic Lieb lattice at the $\Gamma$ and X point are shown in \cref{fig:Lieb}. We indeed find that the amplitudes of quadrupole modes $\psi_2^+$ and $\psi_2^-$ follow the results of the tight-binding model. The weak excitation of dipole and hexapole modes is a result of the inter-orbital coupling that leads to the nontrivial second neighbor coupling.

\section{Perturbations}
\begin{figure}
	\centering
	\includegraphics[width=\textwidth]{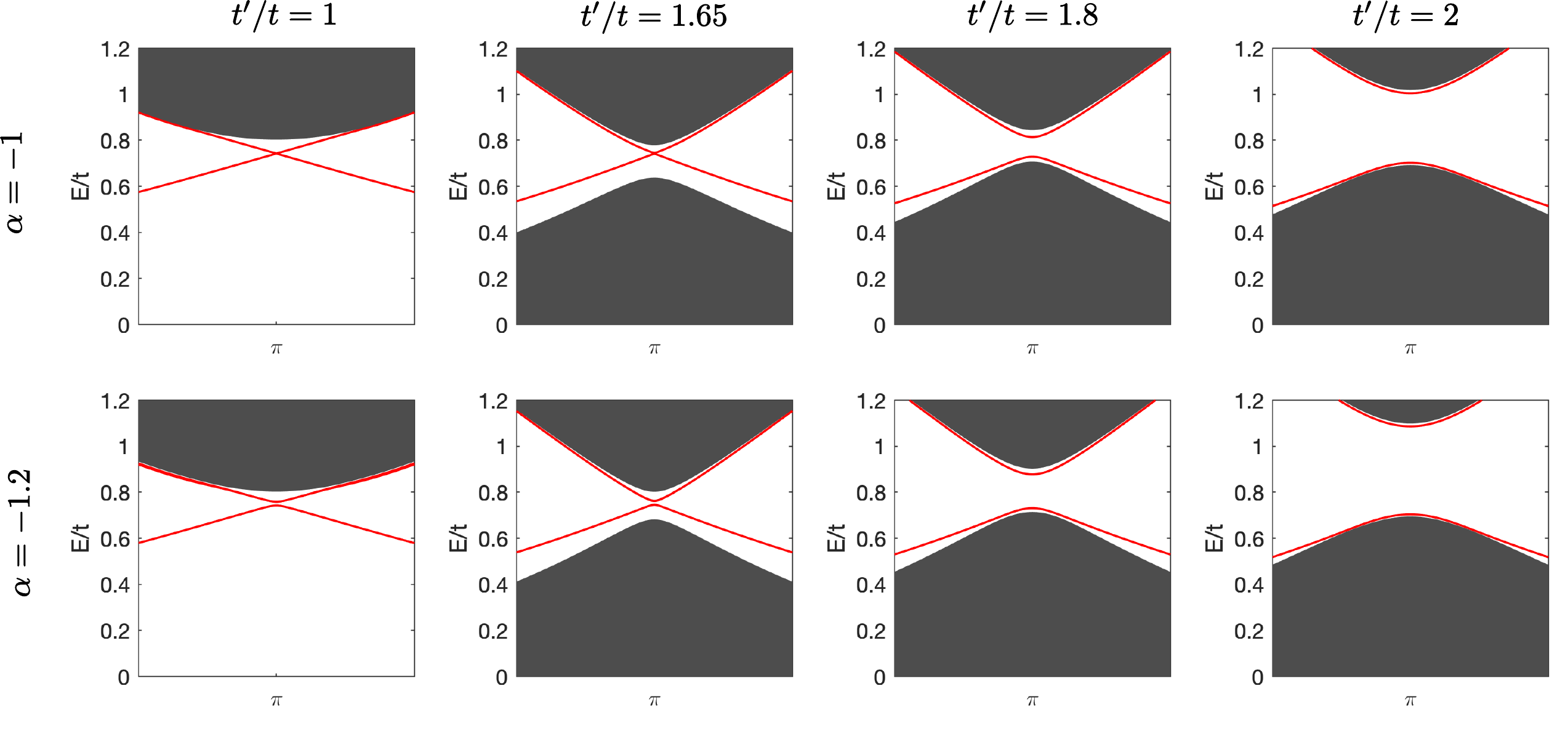}
	\caption{Edge modes from tight-binding model under an inversion breaking perturbation. Top row shows results for the ideal DIII class as a function of the inversion breaking perturbation. Bottom row shows results for the non-ideal $\alpha\neq -1$ case.}
	\label{fig:inversion}
\end{figure}

In this section the inversion breaking perturbation and on-site energy perturbations are discussed using the tight-binding model. Inversion symmetry is broken when the hopping parameters no longer satisfy $t_{22}^{\pm\pm}(\mathbf{a}_i)=t_{22}^{\pm\pm}(-\mathbf{a}_i)$ where $i=x,y$. Define the parameters $t\equiv t_{22}^{++}(\mathbf{a}_x)=t_{22}^{++}(\mathbf{a}_y)$ and $t'\equiv t_{22}^{++}(-\mathbf{a}_x)=t_{22}^{++}(-\mathbf{a}_y)$. Then the intra-orbital Hamiltonian $h_+(\mathbf{k})$ may be written as
\begin{equation*}
	h_+(\mathbf{k}) =
	\begin{pmatrix}
		0 & t+t'e^{-ik_x} & 0 \\
		t+t'e^{ik_x} & 0 & t+t'e^{-ik_y} \\
		0 & t+t'e^{ik_y} & 0
	\end{pmatrix}.
\end{equation*}
The full Hamiltonian including the effective second neighbor coupling then follows from eq.5 of the main text. The strength of the inversion breaking perturbation is quantified using the ratio $t'/t$. Edge modes in the upper gap as a function of $t'/t$ for both $\alpha=-1$ and $\alpha\neq -1$ are shown in \cref{fig:inversion}. The effective second neighbor coupling strength is set to $t_s=0.2t$. We find that for both the ideal and non-ideal cases, the edge modes survive until the gap is closed and reopened. Edge modes in the lower gap undergo the same behavior. 

\begin{figure}
	\centering
	\includegraphics[width=\textwidth]{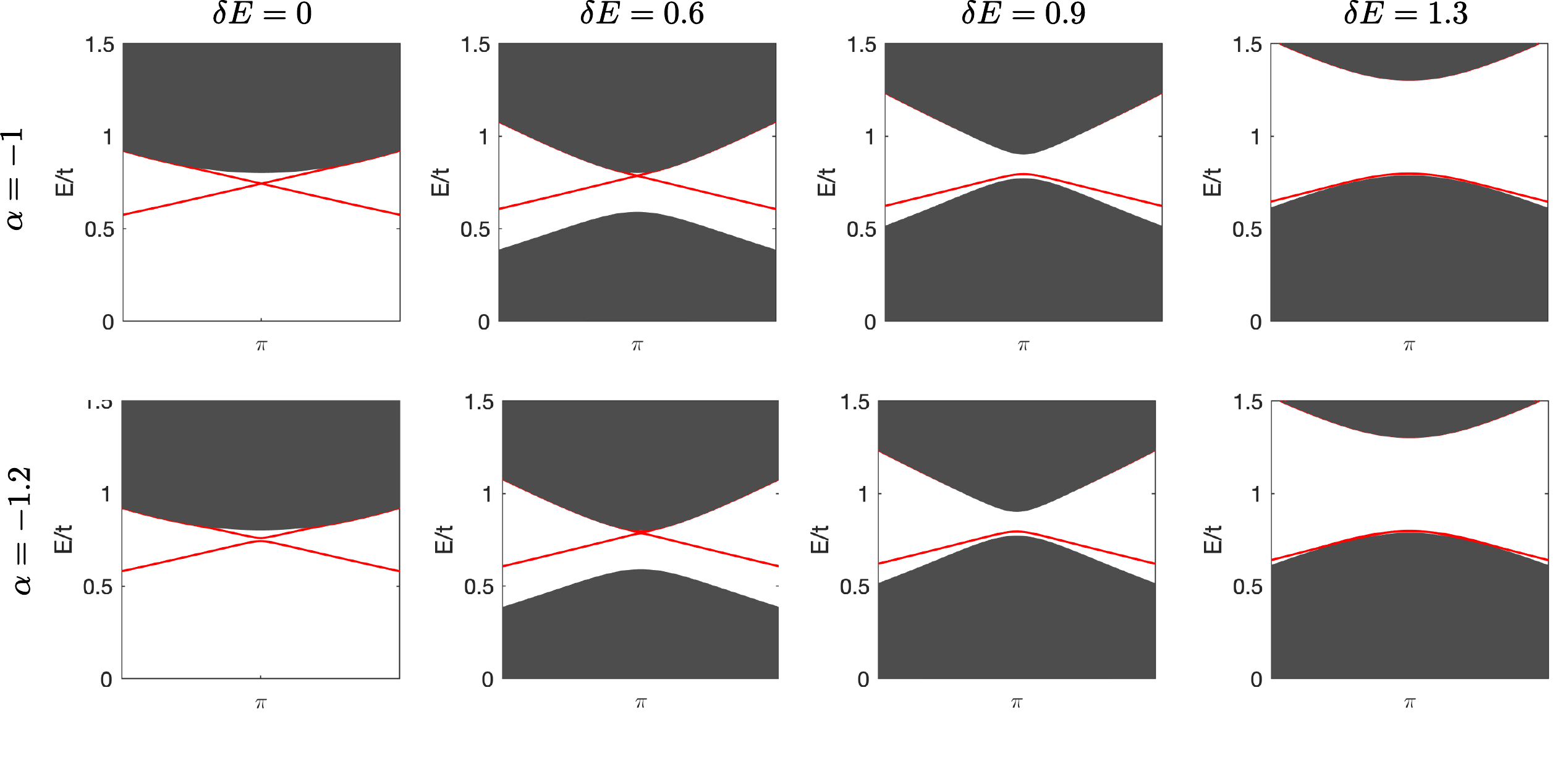}
	\caption{Edge modes from tight-binding model under an on-site energy perturbation. Top row shows results for the ideal DIII class as a function of the perturbation. Bottom row shows results for the non-ideal $\alpha\neq -1$ case.}
	\label{fig:ose}
\end{figure}

An on-site energy is added by modifying the intra-orbital Hamiltonian to 
\begin{equation*}
	h_+(\mathbf{k}) =
	\begin{pmatrix}
		0 & t+te^{-ik_x} & 0 \\
		t+te^{ik_x} & \delta E & t+te^{-ik_y} \\
		0 & t+te^{ik_y} & 0
	\end{pmatrix}.
\end{equation*}
Edge modes in the upper gap as a function of $\delta E$ for both $\alpha=-1$ and $\alpha\neq -1$ are shown in \cref{fig:ose}. Once again we find that the edge modes survive until the gap is closed. Unlike the inversion breaking perturbations, for the on-site energy the gap is closed in only one of the two gaps.

\section{Scaling with size and Fermi energy}
\begin{figure}
	\centering
	\includegraphics[width=\textwidth]{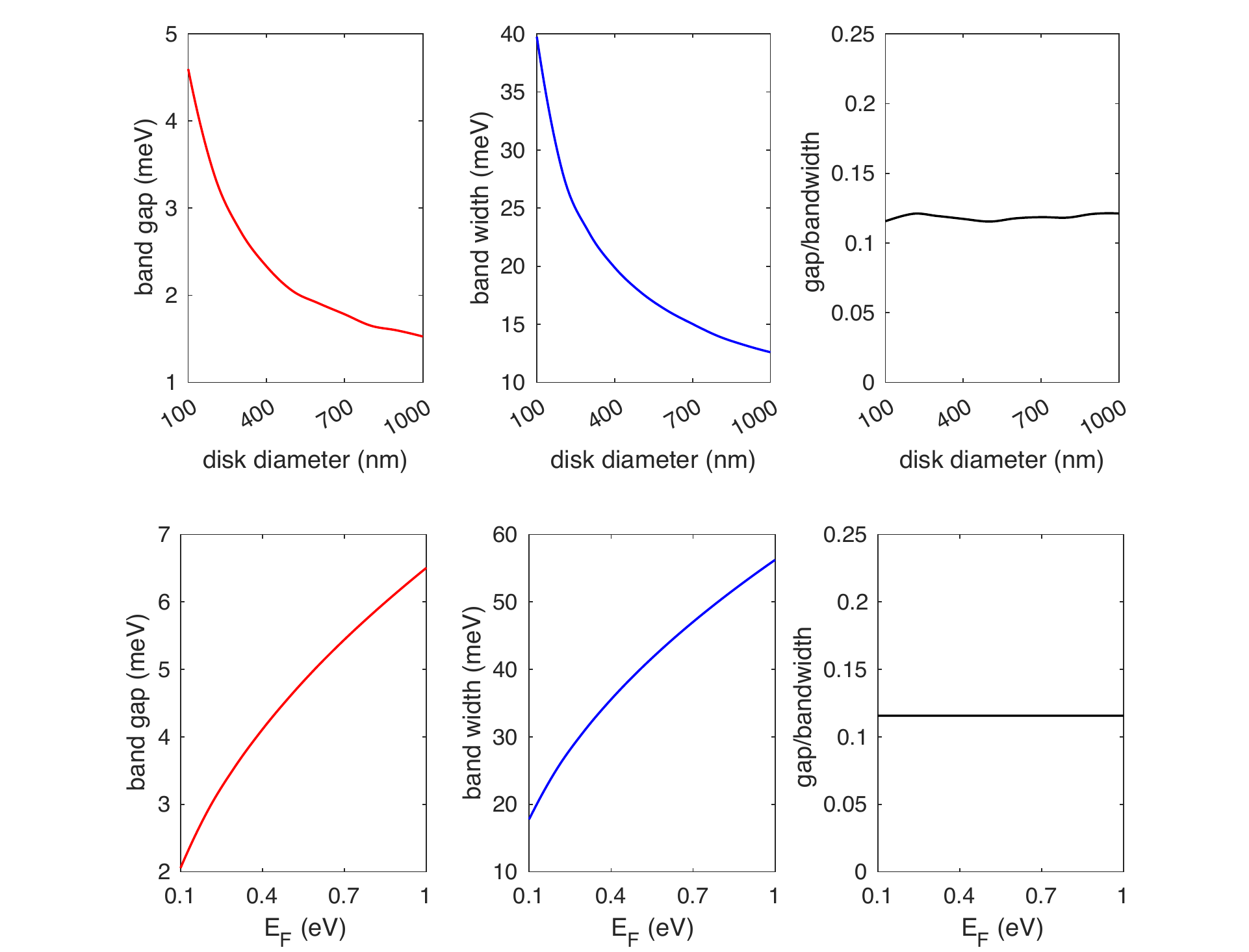}
	\caption{Scaling of band gap and band width of quadrupole band structure. Top row shows scaling as a function of the disk diameter with Fermi energy fixed to 0.5eV. All other dimensions are scaled proportionally. Bottom row shows scaling as a function of Fermi energy with disk diameter fixed to 100nm. }
	\label{fig:scaling}
\end{figure}

Scaling of the quadrupole bands as a function of the geometry dimensions and graphene Fermi energy is examined. The geometry is scaled such that the ratio between the disk size and disk spacing is constant. The Fermi energy is held constant at 0.5eV for the geometry scaling study. We find that both the band width and band gap are reduced as the geometry size is increased (see top row of \cref{fig:scaling}). Interestingly, the ratio between the band width and band gap remains constant even as the geometry is scaled. This implies that the topological physics observed in the system remains unchanged. For the Fermi energy, we find that the band width and band gap increase as the Fermi energy is increased (bottom row of \cref{fig:scaling}). Similarly, the ratio between band width and band gap is unchanged as a function of Fermi energy.

\clearpage
\begin{figure}
	\centering
	\includegraphics[width=0.8\textwidth]{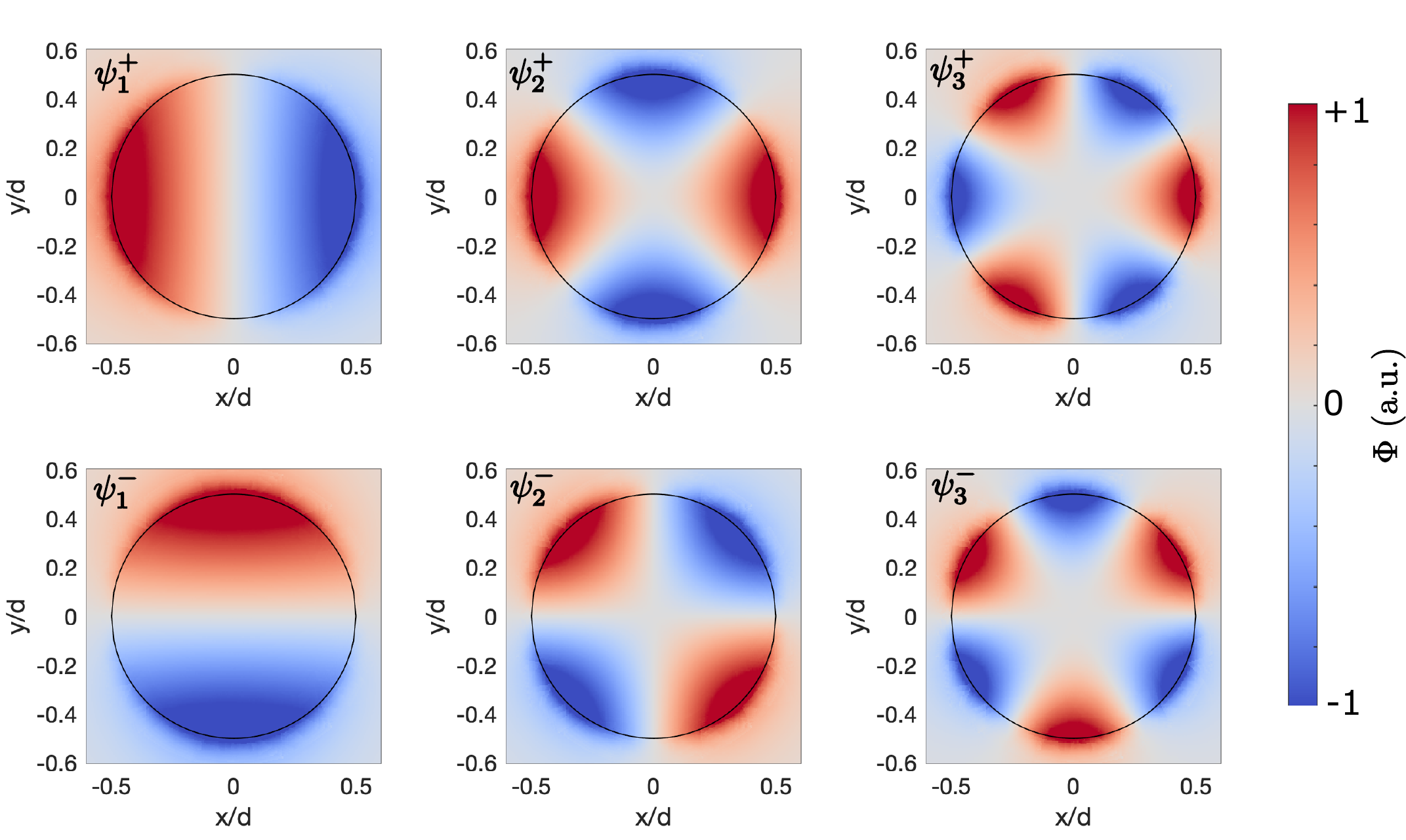}
	\caption{Potential of $\psi_l^p$ eigenstates}
	\label{fig:realbasis}
\end{figure}
